\documentclass[epj,nopacs]{svjour}
\usepackage{graphics}
\begin{document}
\title{A simple Discrete-Element-Model of Brazilian Test}
\author{Sumanta Kundu\inst{1} \thanks{e-mail: sumanta492@gmail.com} 
       \and Anna Stroisz\inst{2} \thanks{e-mail: anna.stroisz@sintef.no} 
       \and Srutarshi Pradhan\inst{2} \thanks{e-mail: srutarshi.pradhan@sintef.no}
}                     
%
%
\institute{Satyendra Nath Bose National Centre for Basic Sciences,
           Block-JD, Sector-III, Salt Lake, Kolkata-700106, India \\
           \and           
           Formation Physics Department, SINTEF Petroleum Research, 
           NO-7465 Trondheim, Norway \\
           }
\date{Received: date / Revised version: date}
%
\abstract{
We present a statistical model which is able to capture some interesting 
features exhibited in the Brazilian test of rock samples. The model is based 
on elements which break irreversibly when the force experienced by the elements 
exceed their own load capacity. If an element breaks the load capacity of the 
neighboring elements are decreased by a certain amount, assuming weakening effect 
around the defected zone. From the model we numerically investigate the stress-strain 
behavior, the strength of the system, how it scales with the system size and also 
it's fluctuation for both uniform and Weibull distribution of breaking thresholds 
in the system. To check the validity of our statistical model we perform few Brazilian 
tests on Sandstone and Chalk samples. The stress-strain curve from model results 
agree qualitatively well with the lab-test data. Also, the damage profile right at 
the point when the stress-strain curve reaches it's maximum is seen to mimic the 
crack patterns observed in our Brazilian test experiments.
} 
\maketitle
\section{Introduction}
\label{intro}

The tensile strength of the rock is one of the important parameters that influences 
the rock crushing and rock blasting results. To measure the tensile strength of a 
solid body (rock, concrete, ceramics etc.) the Brazilian test \cite{Li2013,Cai2004,Erarslan2012,Fahad1996} 
is performed. This test is also named as the Diametral Compression test and the Indirect 
Tensile test. The test was introduced by Carniero \cite{Carneiro1943} in Brazil, and Akazawa
\cite{Akazawa1943} in Japan in $1943$. 

In the Brazilian test a circular disk is diametrically compressed. Due to this compression 
a tensile stress develops perpendicular to the direction of the applied force and the 
amplitude of the induced stress is proportional to the applied force. When the induced 
tensile stress crosses certain limit, fracture develops mainly in the middle zone of the 
sample. The indirect tensile strength of a cylindrical sample with diameter $D$ and thickness 
$t$ is given by 
\cite{Li2013},
\begin{equation}
\sigma_t=\frac{2F}{\pi Dt}
\end{equation} 
where $F$ is the force at the failure point of the material. This equation was obtained
analytically based on the assumption that the rock is isotropic and homogeneous 
\cite{ISRM1978,Simpson2014}. However, the nature is more complex. Therefore we must take 
into account the anisotropy and heterogeneity effects for better estimation of rock-strength.

Besides the tensile strength, the crack initiation and crack propagation upon the 
Brazilian test is often examined \cite{Cai2013,Malan1994}. A lot of studies
have been done on the crack initiation position in the Brazilian test both experimentally
and numerically \cite{Zhu2006}. In this paper, we present our experimental study of 
Brazilian test on two samples and then present a very simple model in the framework of 
statistical physics, which is able to capture some features exhibited in the Brazilian test.

\begin{figure}[t]
\centering
\resizebox{0.40\textwidth}{!}{%
\includegraphics{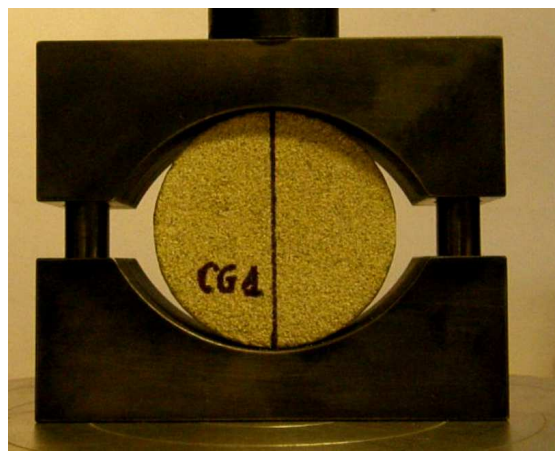}} 
\caption{The loading apparatus used in our experiment.}
\label{fig:1}
\end{figure}
 
 In Sect.~\ref{Sec:2} we describe the experimental procedure of our Brazilian test 
followed by some experimental results in Sect.~\ref{Sec:3}. We elaborately describe 
our model in Sect.~\ref{Sec:4} and present numerical results obtained from this model 
in Sect.~\ref{Sec:5}. Here we investigate stress-strain behavior, damage profile and 
critical strength of the system. Finally we discuss the application of this work and 
make some concluding remarks in Sect.~\ref{Sec:6}.

\section{Experimental Procedure}
\label{Sec:2}
\begin{figure}[t]
\centering
\resizebox{0.40\textwidth}{!}{%
\includegraphics{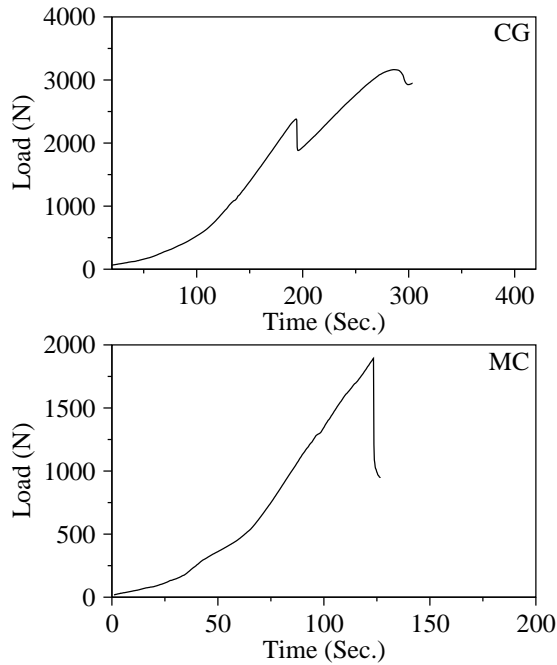}} 
\caption{The load-time behavior of Castlegate sandstone (CG) and Mons chalk (MC) 
samples.}
\label{fig:2}
\end{figure}

The Brazilian (splitting) indirect tensile strength test \cite{Simpson2014} has 
been  performed on two rock types - Castlegate sandstone (CG) and Mons chalk 
(MC). Both rocks have a fairly 
homogeneous and isotropic (with no distinct layering) structure. The specimens were 
cored perpendicular to bedding and cut into the discs of approximately $22$ mm thickness 
and $52$ mm diameter. Each disc was wrapped around the circumference with a masking tape. 
This softened the contact between the rock and the steel curved jaws (Fig. \ref{fig:1}) 
within which the disc is clamped during the test. This curved jaws is then placed in a 
loading frame. We then continuously increase the externally 
applied load, at a very slow but constant speed of crosshead, until the failure of the 
sample occurs. In our experiment the external force is applied with a MTS load frame at the rate 
$0.003$ mm/Sec., until the failure of the material. The reaction force given by the sample is recorded at 
every time. Fracture development, indicated by the sudden drop in loading, is captured using 
a digital camera.

\section{Experimental Results}
\label{Sec:3}

\begin{figure}[t]
\centering
\resizebox{0.40\textwidth}{!}{%
\rotatebox{180}{
\includegraphics{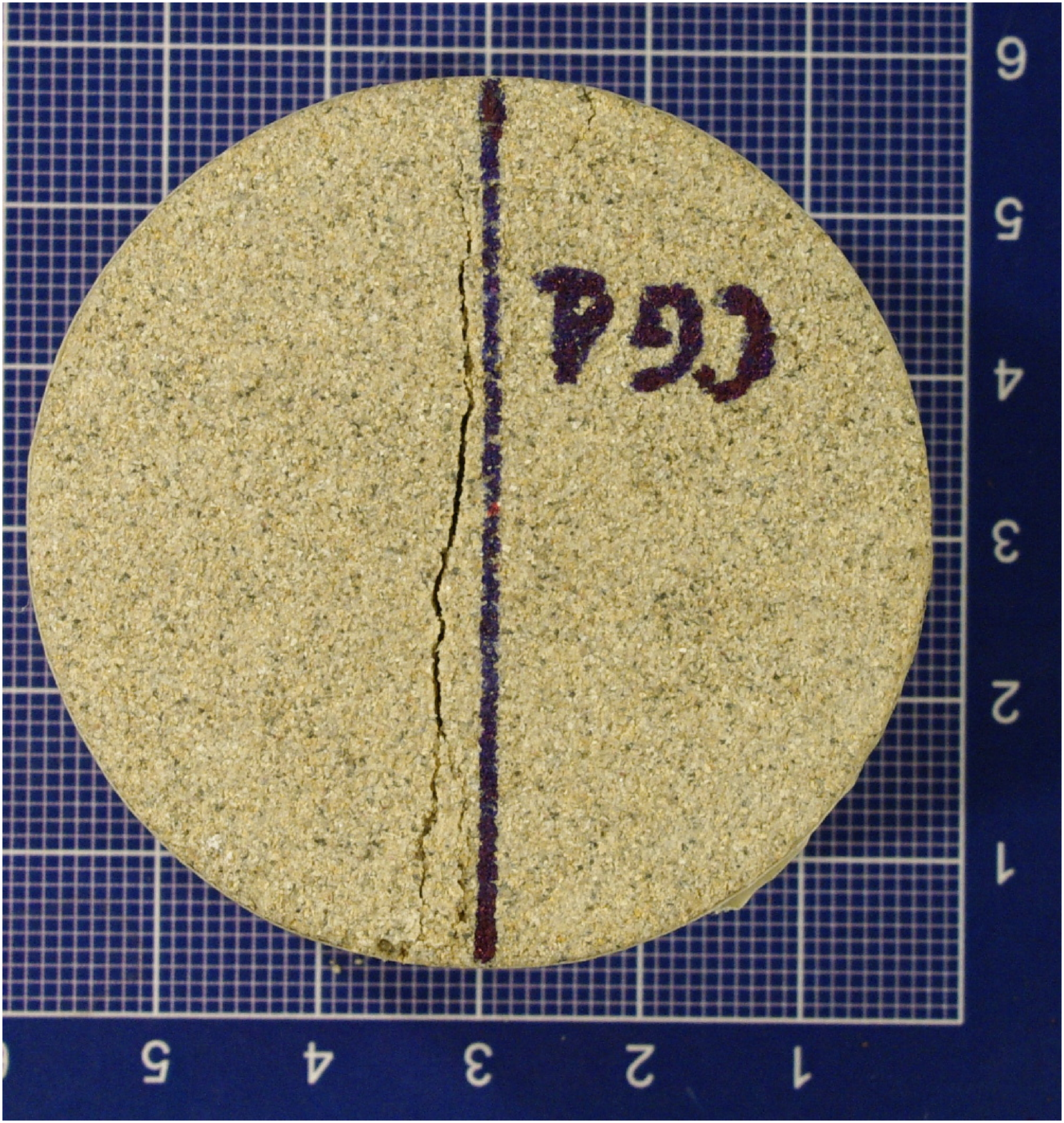}}} \\
\resizebox{0.40\textwidth}{!}{%
\rotatebox{180}{
\includegraphics{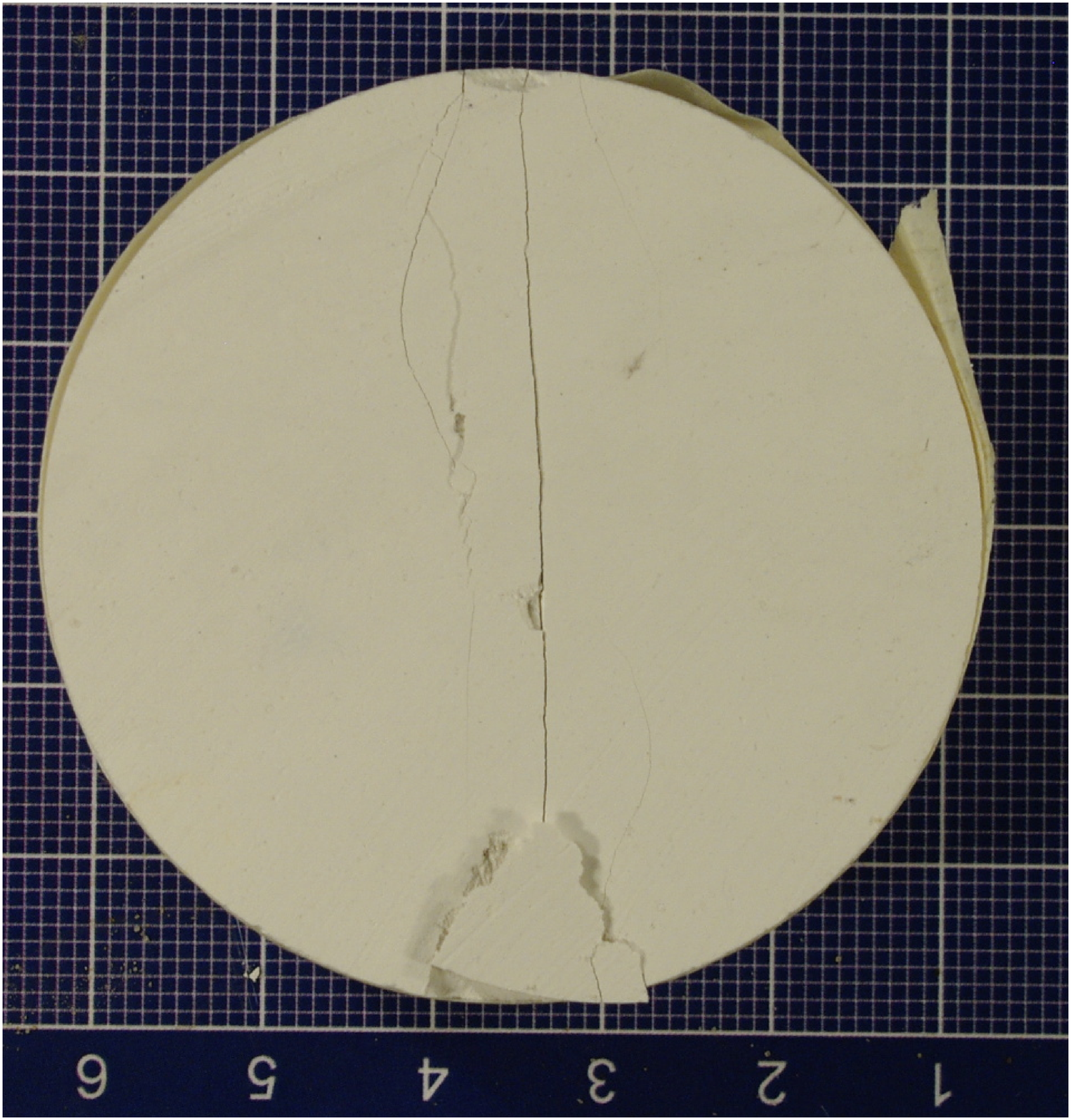}}}
\caption{Fracture patterns after failure. Top one is for Castlegate sandstone (CG) and 
the lower one is for Mons chalk (MC) samples.}
\label{fig:3}
\end{figure}
\begin{table}[b]
\caption{Details about the samples that we examined. The thickness/diameter ratio is 
$0.42$ for both of the samples.}
\begin {tabular}{cccc} \\ 
\hline\noalign{\smallskip}
Sample name & Weight & Diameter & Thickness \\ \hline \hline
Castlegate sandstone     & 91.61gm. & 52.26mm. & 21.78mm. \\
Mons chalk           & 71.73gm. & 51.75mm. & 21.73mm. \\ 
\noalign{\smallskip}\hline
\end {tabular}
\label{tab:1}
\end {table}
\begin{figure}[t]
\centering
\resizebox{0.40\textwidth}{!}{%
\includegraphics{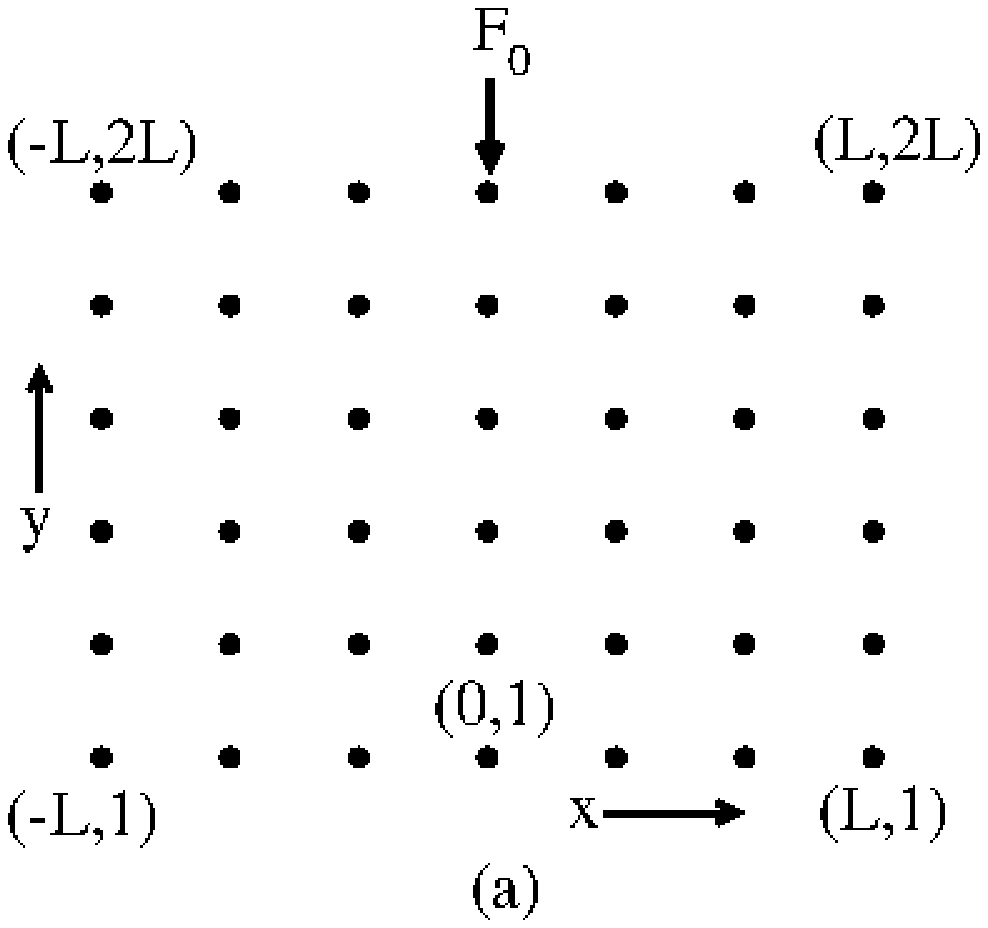}}\\ 
\resizebox{0.40\textwidth}{!}{%
\includegraphics{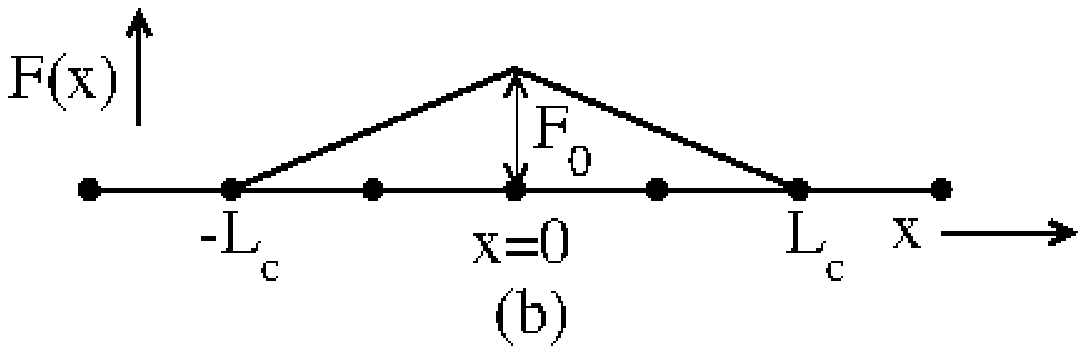}}
\caption{(a) Square lattice of size $L=3$. Each point on the lattice is an element, 
follows Hooke's law before the breaking point. A force $F_0$ is applied at the 
middle column from the top of the lattice. (b) Illustrates the effect of the 
applied force. Here it is assumed that the effect decreases linearly from the 
point of loading. The cut off length $L_c$ and the $F_0$ determines the slope of 
the decreasing force profile, slope $m$=-$F_0/L_c$.}
\label{fig:4}
\end{figure}

 We experimentally examined the strength of two cylindrical samples CG and MC -masses
and dimensions are given in the Table \ref{tab:1}. In our experiment we increase the 
externally applied load on the material and measure the force with time. In Fig. \ref{fig:2} 
the load-time data has been plotted for both samples. The force/load values are measured in
units of Newton (N). We define the tensile strength of material as the maximum load that a 
material can sustain without failure. For CG sample the first drop in the measured load at 
around 195 Sec., signifies the failure. When a fracture opens the near-by sand grains occupy 
the fractured region, also the contact area to the curved jaws broadens and thus the reaction 
force increases further. This is not observed in case of chalk sample due to the brittleness 
of the chalk. From the load-time plot it is obvious that the tensile strength of CG is greater 
than that of MC. In Fig. \ref{fig:3} the damage zone has been shown for both of the samples 
after the failure.       
\begin{figure}[t]
\centering
\resizebox{0.40\textwidth}{!}{%
\includegraphics{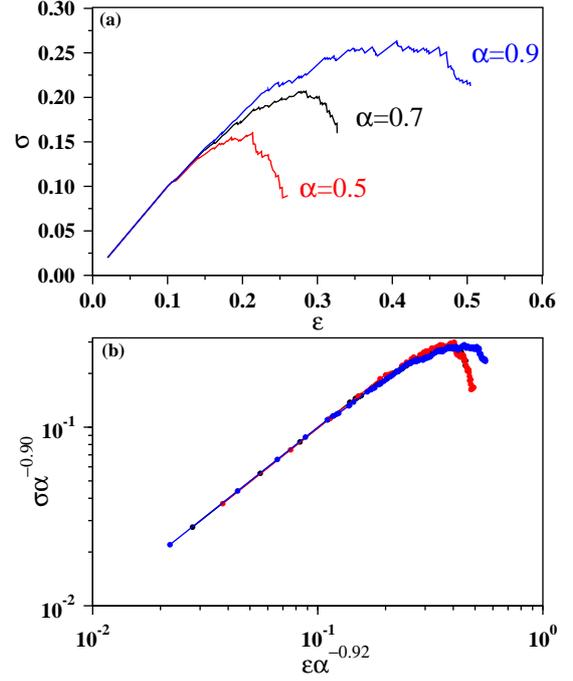}} 
\caption{(a) For the uniformly distributed thresholds between $0.1$ to $1$ and $L_c=L/12$, the 
stress-strain behavior of a system with $L=64$ has been shown for three different values of 
strength decrement factor $\alpha=0.50$ (red), $\alpha=0.70$ (black) and $\alpha=0.90$ (blue). 
(b) The same data in (a) are plotted against the scaled variable $\varepsilon \alpha^{-0.92}$ and
we obtain a reasonably good data collapse upto the failure point (the peak position).
}
\label{fig:5}
\end{figure}

\section{Model}
\label{Sec:4}
\begin{figure}[!t]
\centering
\resizebox{0.40\textwidth}{!}{%
\includegraphics{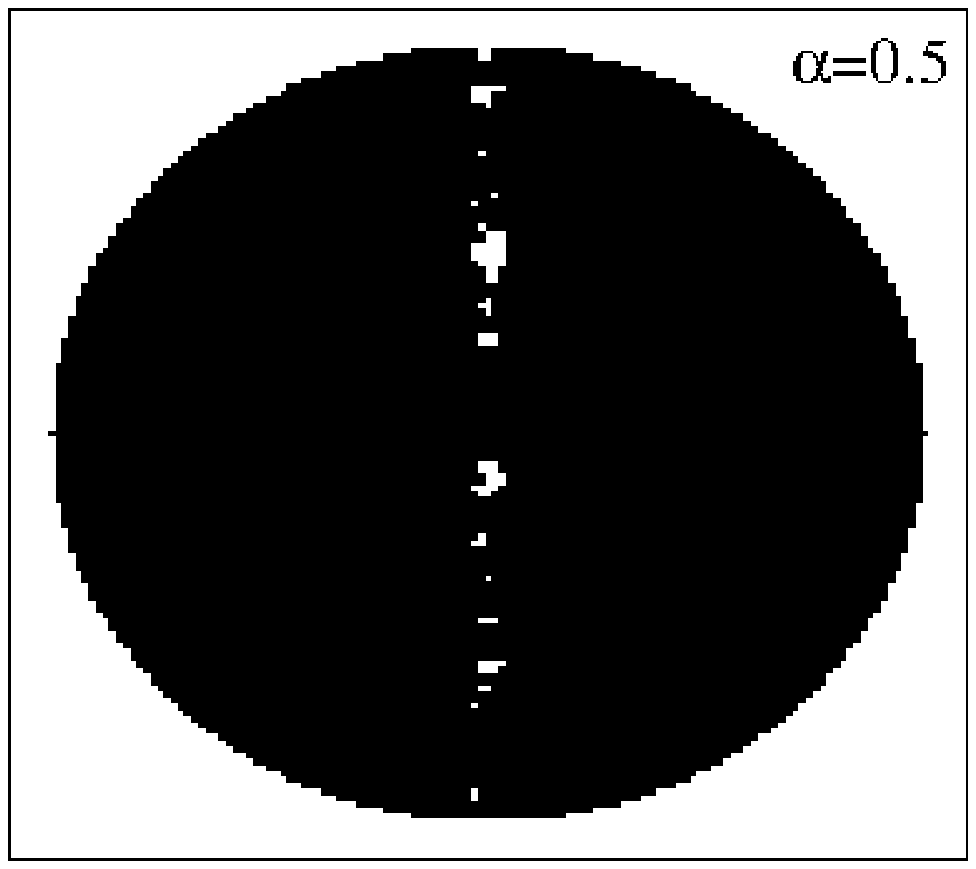}} \\
\resizebox{0.40\textwidth}{!}{%
\includegraphics{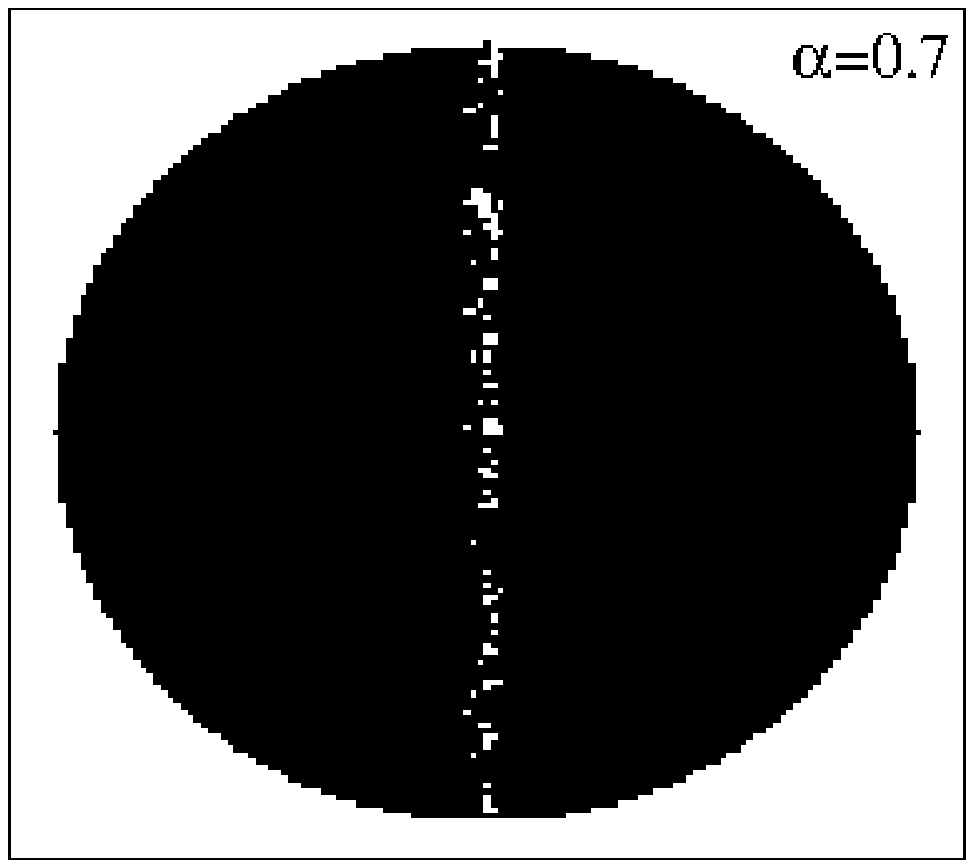}} \\
\resizebox{0.40\textwidth}{!}{%
\includegraphics{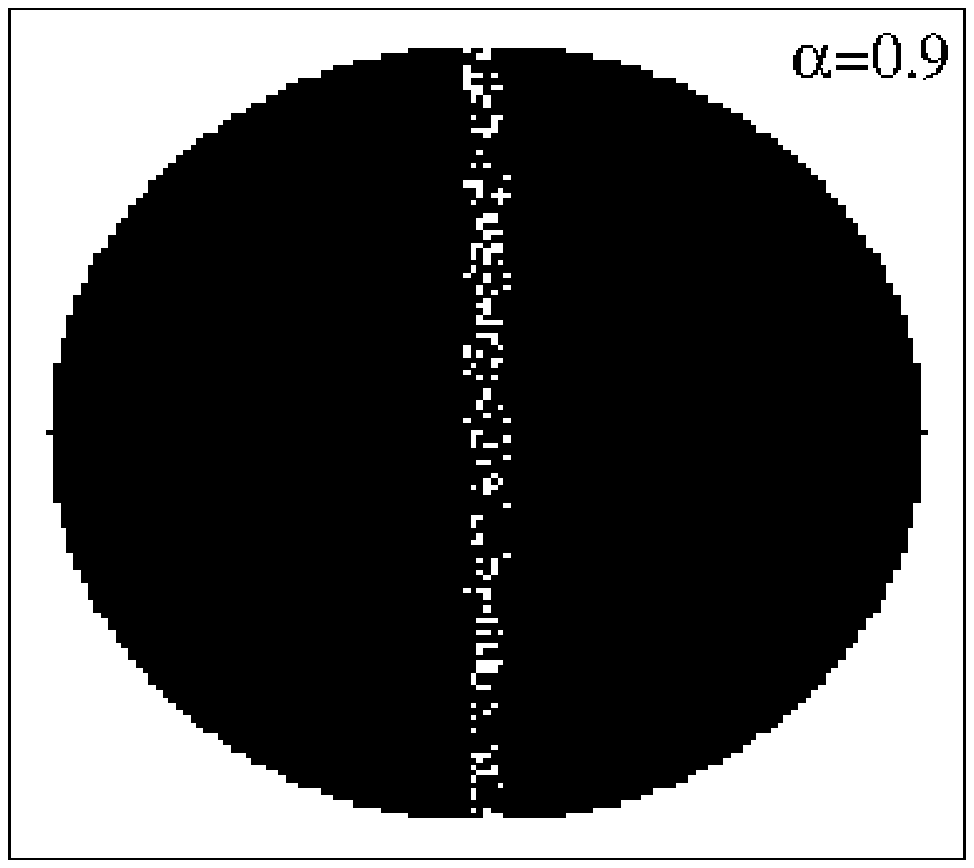}}
\caption{Crack profile for different values of the strength decrement factor, $\alpha$, 
right at the point where the stress-strain curve reached it's 
maximum. The black color represents the intact elements and the white patches at the 
middle of it are the broken elements within the system of size $L=64$. Here the 
thresholds are distributed uniformly between $0.1$ to $1$ and in simulation we used 
$L_c=L/12$.}
\label{fig:6}
\end{figure}

 Our model consists of Hookean elements, placed on each lattice point of a 
regular square lattice, with their own breaking thresholds drawn randomly  
from a given probability distribution. Once the load on an element exceeds 
its breaking threshold, it breaks immediately and the load capacity of its 
neighboring intact elements are decreased indicating the weakening effect 
around the defected zone. One may think of the redistribution of load carried 
by the broken element among the intact elements as in the case of fiber 
bundle model \cite{Hansen2015,Pradhan2010}. In our model the load experienced by the intact 
elements are kept constant throughout the breaking process, elements break only 
due to the decrement of their breaking thresholds. To model the Brazilian test, 
the external load is applied at the middle point of the system and it is assumed 
that the effect of the applied force diminishes linearly as we move-on horizontally 
in either sides from the point of loading. From this model we numerically investigate 
the breaking properties of the system.     

 We consider a square lattice of size $L$ as shown in the Fig. \ref{fig:4}(a). 
Each point on the lattice is an element. Each individual element positioned at 
$(x,y)$ has been assigned a breaking threshold $b_{(x,y)}$  i.e., it can sustain 
a maximum $b_{(x,y)}$ amount of load through it, beyond which it breaks irreversibly. 
Before breakdown each element follows Hooke's law. For simplicity we assume the 
stiffness factor of all the elements to be unity. The breaking thresholds 
$\lbrace b_{(x,y)} \rbrace$ are drawn from a probability distribution $p(b)$, the 
cumulative distribution of which is given by $P(b)=\int_0^b p(y)dy$. The breaking 
threshold distribution is the only source of disorder present in this model.    

 In our model we apply a force at the middle point of the lattice 
from the top of it, as shown in the Fig. \ref{fig:4}(a). We assume that the effect 
of this applied force extends within a certain region from the point of application 
of the force. We consider the effect of this force decreases linearly and this force 
profile is symmetric about the point of loading. The force profile at the positive 
part of the lattice has the form:
\begin{equation}
F(x)=mx+F_0  
\end{equation}   
and since it is symmetric about the point of loading so we have $F($-$x)=F(x)$. 
In Eqn. (2) $F_0$ is the amplitude of the applied force at the point $x=0$ and 
$m$ defines how far the effect of this applied force should extend. In our model 
we prefix a cut-off length $L_c$ beyond which the applied force has no effect i.e., 
$F(x)=0$ for $x > L_c$. This actually defines the value of $m($=-$F_0/L_c)$. The 
whole schematic picture is depicted in the Fig. 4(b). We also assume that the load 
through all the elements of a particular column is same i.e., $F(x,y)=F(x)$.

An element breaks irreversibly and it's neighborhood gets affected, when the force 
through the element exceeds it's breaking threshold. When an element breaks, the 
breaking strength of it's surviving nearest neighbors is decreased by a factor 
$\alpha (0\leq \alpha \leq 1)$, i.e., their new thresholds are assigned a value 
equal to $\alpha$ times of their own previous thresholds. This is actually the sense 
of weakening around the defected zone. The force profile remains unaltered in this model. 

When a force is applied to the system, the breakdown of the elements occur from the 
middle part (close to the $x=0$) of the system since the force field is much stronger 
in this region. So, the local failures will grow from the middle zone of the system. 
This local failures decrease the strength of the elements around the defected zone. 
A local failure may occurs due to the weakness of the elements as well as due to the 
decrement of the strength of the elements after an failure event and due to this a 
spatial correlation  develops in the failure events. Since the force field decreases 
linearly in the horizontal direction, it is more probable that failure will occur in 
the vertical direction. This makes cracks to propagate in the vertical up or vertical 
down direction rather than horizontal direction from the defected zone, which correlates 
with the experimental observations.       

The width of disorder, the value of the cut-off length $L_c$ and the strength decrement 
factor $\alpha$ play a crucial role in the crack propagation. For a fixed window of disorder, 
smaller the value of $\alpha$ and $L_c$, lead successive failure to be more and more localized 
and directed in the vertical direction.

\section{Numerical Results}
\label{Sec:5}

We investigate the stress-strain behavior of the system which is the most crucial part 
of the Brazilian test. We observe the stress-strain behavior, the system size dependence 
of the maximum value of the stress and it's fluctuation for both uniform and Weibull
distribution of the breaking thresholds.

To obtain the stress-strain curve we increase the external load quasi-statically. Due 
to the loading all elements get stretched. At an elongation, $\varepsilon$ per intact 
element, the total force at a column is $\varepsilon$ times the total number of intact 
elements at the column. We define stress ($\sigma$) as-
\begin{equation}
\sigma=\frac{N_{in}\varepsilon}{2L}
\end{equation} 
where, $N_{in}$ is the number of intact elements at the central column.
\begin{figure}[t]
\centering
\resizebox{0.40\textwidth}{!}{%
\includegraphics{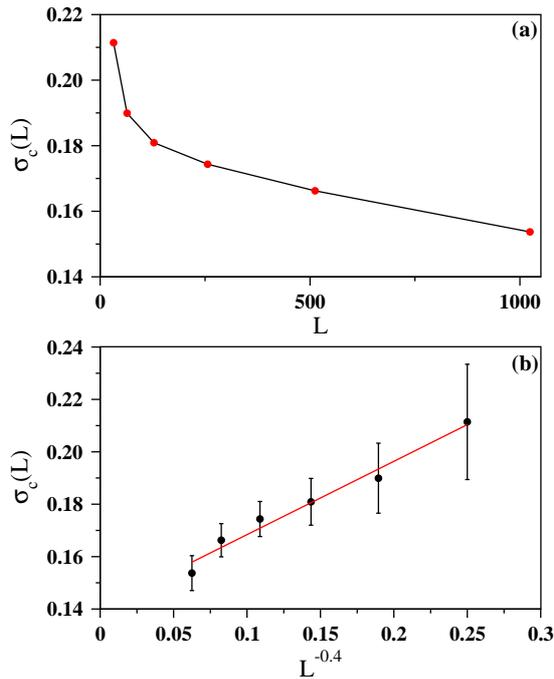}} 
\caption{(a) Plot of $\sigma_c(L)$ with $L$. The system size varies from $L=32$ to $L=1024$ 
with an increment of factor $2$. (b) Plot of $\sigma_c(L)$ with $L^{-0.4}$ is best fitted by 
a straight line, the extrapolation of which gives the value of $\sigma_c=0.14$.}
\label{fig:7}
\end{figure}

Numerically, the stress-strain curve is obtained in the following way. Initially we 
start with a fully intact system and then we apply a load such that the weakest element 
in the system breaks. The weakest surviving element is determined by calculating the 
maximum of the ratio between force field to the breaking threshold i.e., 
$max\lbrace F(x,y)/b_{(x,y)} \rbrace$, which is equal to $max\lbrace F(x)/b_{(x,y)} \rbrace$. 
Once it is removed, the breaking strength of it's nearest surviving neighbor is reduced 
by a factor $\alpha$. Due to this, it may so happen that the load through some of the 
elements exceed their breaking threshold and we break them simultaneously again reducing 
the strength of their neighboring elements, that may cause further breakdown and so on. 
This process stops when all the remaining intact elements have their breaking strength 
greater than their respective experienced stress values. We then increase the externally 
applied load in such a way that the weakest element among the remaining intact system reaches 
it's breaking threshold and then the entire procedure is repeated. This procedure is continued 
until we are at a point well above the point where the maximum of the stress-strain curve takes 
place. 

In Fig. \ref{fig:5}(a) the stress-strain curve is shown for a particular configuration 
for different values of $\alpha$. In the next figure by suitably scale the abscissa and ordinate 
we could obtain a data collapse (Fig. \ref{fig:5}(b)) upto the peak point. In experiments, after 
peak position, different rock samples behave differently -depending on their failure structure 
(clear fracture, complex fracture, crashed etc). Therefore, in our numerical study, we have tried 
the scaling up to the peak position. The defected zone right at the point where the stress-strain 
curve reaches it's maximum has been shown by the white color in Fig. \ref{fig:6}. Throughout the 
simulation we have used $L_c=L/12$.

Once we generate the stress-strain curve, we then calculate the critical strength of 
the system. We define critical strength $\sigma_c^s(L)$ for a particular sample 
$s$ of size $L$ with a particular set of breaking thresholds $\lbrace b_{(x,y)} \rbrace$ 
as the maximum value of the stress in the stress-strain curve. For each sample $s$ we 
numerically calculate $\sigma_c^s(L)$ and when the whole procedure is repeated over 
a large number of uncorrelated samples, it eventually gives the average value of the 
critical strength $\sigma_c(L) = \langle \sigma_c^s(L) \rangle$ for the system of 
size $L$. We then repeat the entire set of calculation for different values of $L$ to 
know how $\sigma_c(L)$ approaches to it's asymptotic value. 

\begin{figure}[t]
\centering
\resizebox{0.40\textwidth}{!}{%
\includegraphics{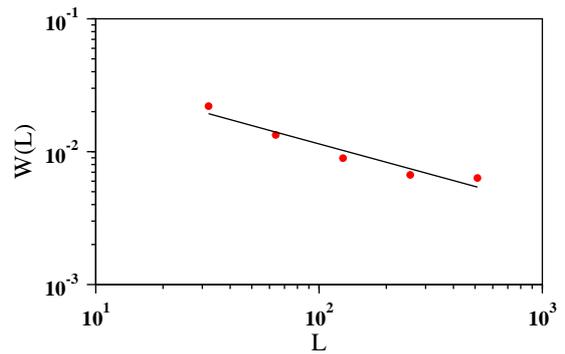}} 
\caption{Log-log plot of $W(L)$ with $L$ for system size upto $L=512$. The fitted line has a
slope $0.46$ and thus we conclude $W(L) \sim L^{-0.46}$.}
\label{fig:8}
\end{figure}
\begin{figure}[t]
\centering
\resizebox{0.40\textwidth}{!}{%
\includegraphics{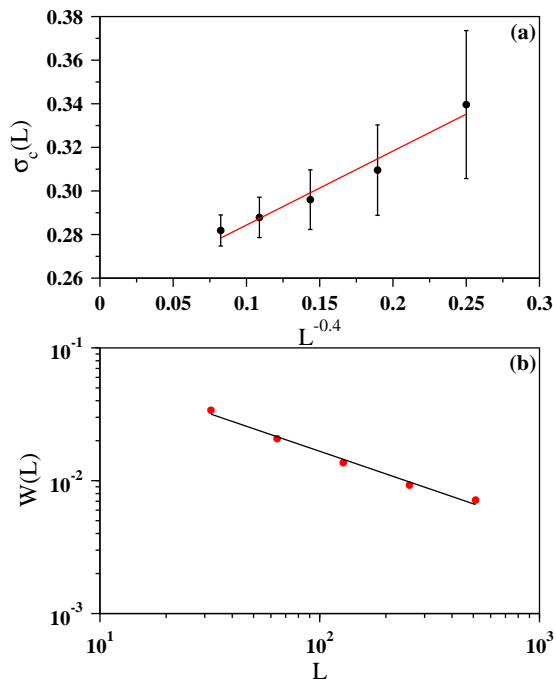}} 
\caption{(a) Plot of $\sigma_c(L)$ with $L^{-0.4}$. The best fitted straight line 
when extrapolated, it cuts the vertical axis at $0.24$ and thus the value of $\sigma_c=0.24$. 
Here the breaking threshold of the elements are drawn from the Weibull distribution of the 
form of Eqn. (6). (b) The fluctuation $W(L)$ varies with $L$ as- $W(L) \sim L^{-0.51}$.}
\label{fig:9}
\end{figure}

We assume that the average value $\sigma_c(L)$ for a system of size $L$ converges to it's 
asymptotic value $\sigma_c = \sigma_c(\infty)$ as $L\rightarrow \infty$ according to the 
following form,
\begin{equation}
\sigma_c(L) = \sigma_c+AL^{-1/\nu}
\end{equation}
For uniformly distributed breaking threshold of the elements in the range [$0.1$-$1.0$], the 
variation of $\sigma_c(L)$ against $L$ has been shown in Fig. \ref{fig:7}(a). When the same 
$\sigma_c(L)$ values are plotted in Fig. \ref{fig:7}(b) as a function of $L^{-1/\nu}$ with $1/\nu=0.40$, 
we observe that it can be fitted by a straight line. The least-squares fitted straight line 
cuts the vertical axis and has the form $\sigma_c(L) = 0.14+0.28L^{-1/\nu}$. From here our 
conclusion is that $\sigma_c=0.14$ and $\nu=5/2$ for the uniformly distributed breaking 
thresholds. 

We also measure the fluctuation of the critical strength, defined as- 
\begin{equation}
W(L)=\sqrt{\langle \sigma_c(L)^2 \rangle-\langle \sigma_c(L) \rangle^2}
\end{equation} 
In Fig. \ref{fig:8} we plot $W(L)$ for different values of $L$ in log-log scale for the 
uniformly distributed breaking threshold and we observe that it decreases with increase 
in $L$. The best fitted line has a slope $0.46$. 

We then repeat our entire set of studies for elements with Weibull breaking threshold 
distribution of the form:
\begin{equation}
P(b)=1-e^{-b^2}
\end{equation} 
We have estimated the values of $\sigma_c(L)$ numerically for six different system sizes from 
$32$ to $1024$ increased by a factor of $2$ at every step. Plotting them against $L^{-0.4}$ 
and extrapolating as $L\rightarrow \infty$ we have obtained $\sigma_c=0.24$. In Fig. 
\ref{fig:9}(a) the plot of $\sigma_c(L)$ with $L^{-0.4}$ has been shown. Our conclusion is 
that for the Weibull distribution of breaking thresholds of the form of Eqn. (6), the value 
of $\sigma_c=0.24$ and $\nu=5/2$. 

We also calculate $W(L)$ for different values of $L$ upto $L=512$. In Fig. \ref{fig:9}(b) 
we plot the $W(L)$ as a function of $L$ and this is fitted with a straight line, resulting 
$W(L) \sim L^{-0.51}$.

\section{Discussions and Conclusions}
\label{Sec:6}

To model the Brazilian test we have introduced a statistical model and studied
numerically the strength of the material. The stress-strain behavior of the system is
found to have a maxima. We have investigated the critical 
strength of the system and also the system size dependence of the critical strength 
for both uniformly and Weibull distributed breaking threshold of the elements in the 
system. We observe that the damage profile right at the critical point (Fig. \ref{fig:6}) 
resembles the nature of the damage profile observed in the Brazilian test 
experiment. 

In our experiment, we have used the Brazilian test set up as indicated in Fig. 1.
How the force field varies from the point of loading is entirely dependent on the shape
and the material of the loading platen. For simplicity, we assume that the force decreases
linearly from the tip point. It may also be possible to design loading plates where the
gradient of the applied force follows power law or exponential. In our simulation we can
easily include this type of force gradient along the loading plate. We believe that, it
will have little effect on the nature of the final fracture pattern - because the main
fracture initiates at and around the most stressed elements and after initial failure the
newly created weak zones guide the fracture development. Therefore, as long as we have a
force distribution on the system having a stress-tip and symmetrical gradient (decreasing)
at both sides from the tip, we can expect similar fracture pattern.

For breaking thresholds distributed uniformly and with Weibull distribution, we
observe that the critical strength $\sigma_c(L)$ of a system of size $L$ approaches 
the asymptotic values  $0.14$ and $0.24$ respectively. For all the results obtained 
from our model, we have taken the strength decrement factor $\alpha=0.7$ and the 
cut-off length of the applied force $L_c=L/12$.

\begin{figure}[t]
\centering
\resizebox{0.40\textwidth}{!}{%
\includegraphics{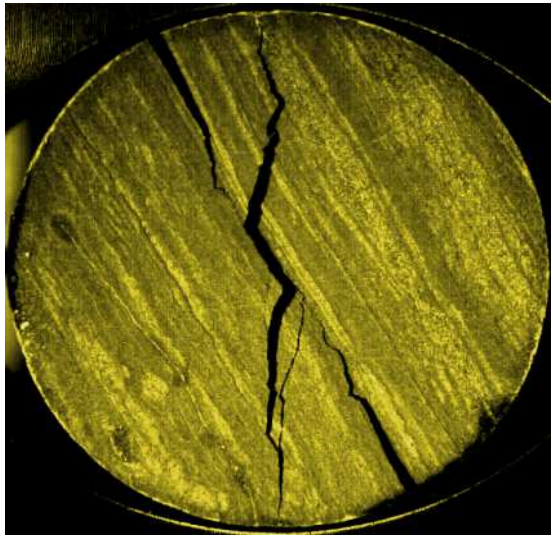}} 
\caption{Snapshot of Brazilian test on an-isotropic shale sample at SINTEF lab 
\cite{Simpson2014}}
\label{fig:10}
\end{figure}

The stress-strain curve is very sensitive to the value of $\alpha$. In Fig. \ref{fig:5} the 
stress-strain behavior of a system with $L=64$ has been shown for three different 
values of $\alpha$. Lowering the value of $\alpha$ the more and more narrow, 
vertically directed and localized the damage zone become, shown in Fig. \ref{fig:6}. As an 
effect the strength of the system will decrease. For $\alpha=1$ the behavior of this 
model is same as the mean field version of the Fiber Bundle model \cite{Hansen2015,Pradhan2010,Roy2013}, 
the elements break one by one with increasing sequence of their breaking thresholds. 

Our statistical model is very simple in the sense -there are only two tuning 
parameters: the strength decrement factor $\alpha$ and the slope of the applied 
force/ load. According to the classical theory -stress fields are enhanced 
around the failure/ defect zones -which in turn weaken the elements in those 
areas. This fact has been captured in our model through the factor $\alpha$.
The second parameter can be treated as an initial boundary condition 
which depends on the geometrical arrangements for the Brazilian test. 
  
Recently Simpson et. al. has done a very interesting Brazilian test 
\cite{Simpson2014} on anisotropic rock sample (shale) and has monitored 
fracture development through high speed camera (See Fig. \ref{fig:10}). Can we 
apply our model to simulate/explain such Brazilian test on highly 
anisotropic rocks \cite{Barla1973}? The answer is yes. We can easily create weak bedding 
planes inside the model rock mass  and we expect that in our model when 
the growing-fracture meets a weak plane - it must follow the direction 
of the weak plane unless it gets arrested by the presence of some strong 
elements. However we have to create the 3D version of our model first and 
then it needs a proper calibration against lab-experiments  -which is our 
ongoing research activity. 

The research work in this paper is a part of the scientific collaboration in 
the INDNOR project 217413/E20 funded by the Research Council of Norway. S. K. 
is thankful for financial support through the project grant to visit SINTEF 
Petroleum Research. We thank S. S. Manna for useful discussions. 
S.P. acknowledges partial support from SINTEF Petroleum Research through 
internal SIP project (7020606). 

SK and SP implemented the model and performed the numerical analysis. AS set-up
the experiment and performed the Brazilian test in presence of other two co-authors.
All authors took part in developing the model, discussing the results and writing
the paper.     
 

\end{document}